\documentclass[12pt,a4paper]{article}
\usepackage{graphicx,amsfonts,amsmath}
\def\Pom{{\bf I\!P}}
\def\Ode{{\bf \!O}}

\newcommand{\vecc}[1]{\mathbf{#1}}
\newcommand{\vecs}[1]{\mathbf{#1}{\lower-.2em\hbox{}^{2}}}

\newcommand{\nn}{\nonumber}

\begin{document}

\title{ Single-spin asymmetry in pion production \\
in polarized proton-proton collisions and odderon }
\author{A.~Ahmedov$^1$, E.~N.~Antonov$^2$,
E.~Barto\v{s}$^1$\footnote{On leave of absence from Comenius
University, 84248 Bratislava, Slovakia}\hspace{2mm},\\
E.~A.~Kuraev$^1$ and E.~Zemlyanaya$^1$ }
\date{}
\maketitle
\begin{center}
$^{1}$ \it Join Institute for Nuclear Research, 141980 Dubna, Russia,\\
$^{2}$ \it Leningrad Institute for Nuclear Physics, Gatchina,
Russia
\end{center}

\begin{abstract}
Single-spin asymmetry appears due to the interference of single
and double gluon exchange between protons. A heavy fermion model
is used to describe the jet production in the interaction of gluon
with the proton implying the further averaging over its mass. As
usually in one-spin correlations, the imaginary part of the double
gluon exchange amplitude play the relevant role. The asymmetry in
the inclusive set-up with the pion tagged in the fragmentation
region of the polarized proton does not depend on the center of
mass energy in the limits of its large values. The lowest order
radiative corrections to the polarized and unpolarized
contributions to the differential cross sections are calculated in
the leading logarithmical approximation. In general, a coefficient
at logarithm of the ratio of cms energy to the pion mass depends
on transversal momentum of the pion. This ratio of the lowest
order contribution to the asymmetry may be interpreted as the
partial contribution to the odderon intercept. The ratio of the
relevant contributions in the unpolarized case can be associated
with the partial contribution to the pomeron intercept. The
numerical results given for the model describe the jet as a heavy
fermion decay fragments.
\end{abstract}

\section{Introduction}
Let us consider the inclusive process of the pion creation in the
fragmentation region of polarized proton at high energy
proton-proton collisions
\begin{equation}
P_1(p_1,a) + P_2(p_2)\to \pi(p) +
X_1(p_1') + X_2(p_2'),
\end{equation}
where $a$ is the transversal to beam (implied by cms) axes spin of
initial proton
\begin{gather}
a=(0,0,\vecc{a}),\quad p=(E\beta,E\beta,\vecc{p}),\quad
E=\sqrt{s}/2,\quad
s=(p_1+p_2)^2\gg \vecc{p}^2\sim m^2, \nn\\
p_1=E(1,\beta_0,0,0),\quad p_2=E(1,-\beta_0,0,0),\quad
\beta_0=\sqrt{1-\frac{m^2}{E^2}},
\end{gather}
where $\beta \sim 1$ is the energy fraction of pion,
$M_{1,2}=\sqrt{p_{1,2}^{'2}}$ are the invariant masses of the
jets, which we will assume to be of the order of nucleon mass $m$.
We study the two jet kinematics with jets $X_1$, $X_2$ moving
along the initial hadron directions. The jet created by the
transversely polarized proton is supposed to contain the detected
pion. Moreover, we consider the case when its production is not
related with the creation of nucleon resonances . In terms of pion
transverse components it corresponds to the condition
\begin{gather}
\tilde{s}_1=(p+p_1')^2=\frac{1}{\beta\bar{\beta}}\left[\beta
M_1^2+(\vecc{p}+\beta\vecc{k_1})^2\right] > M_{res}^2-m^2,\\ \nn
\vecc{p}+\vecc{k_1}+\vecc{p_1'}=0,
\end{gather}
where $\vecc{k_1}$ is the transfer momentum between protons.
Through this paper we use Sudakov parameterization of 4--momenta
of the problem
\begin{equation}
k_i=\alpha_i q_2+\beta_i q_1+k_{i\bot},\quad
k_{i\bot}=(0,0,\vecc{k_i}),\quad q_{1,2}=E(1,\pm 1,0,0).
\end{equation}
\begin{figure}[tb]
\begin{center}
\includegraphics[scale=.8]{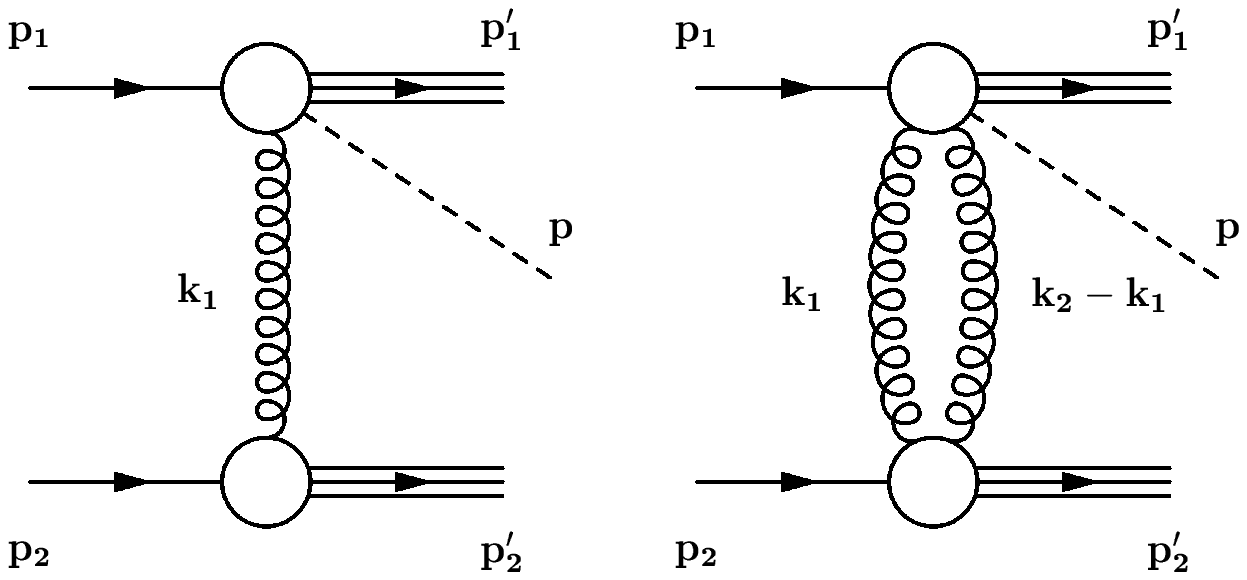}
\caption{One and two gluon exchange in peripheral collision of
protons with $\pi$ production $P_1(p_1,a) + P_2(p_2)\to \pi(p) +
X_1(p_1') + X_2(p_2')$.} \label{fig:1}
\end{center}
\end{figure}
The azimuthal one-spin asymmetry arises from the interference of
the amplitudes with one and two gluon exchanges between nucleons
(see Fig.~1)
\begin{gather} \label{eq:5}
A=\frac{2E_\pi\displaystyle\frac{d^3{\sigma(\vecc{a},\vecc{p})}}
{d^3p}-2E_\pi\frac{d^3 {\sigma(-\vecc{a},\vecc{p})}}{d^3 p}}
{2E_\pi \displaystyle\frac{d^3{\sigma(\vecc{a},\vecc{p})}} {d^3
p}+2E_\pi \frac{d^3\displaystyle{\sigma(-\vecc{a}, \vecc{p})}}{d^3
p}} =\alpha_s R f(\rho,z),\quad R=|\vecc{a}|\sin\varphi,\\\nn
f(\rho,z)= \frac{I_0^{(3)}T_0 +z I_1^{(3)}T_1}{J_0^{(2)}T_2 + z
J_1^{(2)}T_3},\quad \rho = \frac{|\vecc{p}|}{m},\quad
z=\frac{\alpha_s}{\pi}\ln\frac{s}{m^2}
\end{gather}
where $\alpha_s$ is the strong coupling constant, $\varphi$ is the
azimuthal angle between the 2--vectors $\vecc{a}$ and $\vecc{p}$,
transverse to the beam axis. The functions $I$, $J$ as well as the
color factors $T_i$ will be specified below. For convenience we
put here the alternative form for the phase volume of pion
\begin{equation}
\frac{d^3 p}{2E_\pi}=m^2\rho d\rho d\phi\frac{d\beta}{2\beta}.
\end{equation}

The one gluon exchange matrix element has a form
\begin{equation}M_0=i4\pi\alpha_s \frac{{J^{(1a)}_\mu
J^{(2a)}_\nu} g_{\mu\nu}}{{k_1^2}},
\end{equation}
the currents $J^{(1a)}$, $J^{(2a)}$ are associated with the jets
created by particles 1, 2 and index $a$ describes the color state
of the jet. Using Gribov representation of the metric tensor
\begin{eqnarray}
g_{\mu\nu} \approx \frac{2}{s}p_2^\mu p_1^\nu,
\end{eqnarray}
and the gauge condition for the currents
\begin{eqnarray*}
k_1^\mu J^{(1a)}_\mu = (\alpha_1 p_2 + k_{1\bot})J^{(1a)} = 0,\quad
k_1^\mu J^{(2a)}_\mu = (\beta_1 p_1 + k_{1\bot})J^{(2a)} = 0,
\end{eqnarray*}
we express $M_0$ in the form
\begin{eqnarray}
M_0=i8s\pi\alpha_s\sum_{a=1}^{N^2-1}\frac{\vecc{J^{(1a)}}\cdot\vecc{k_1}
\vecc{J^{(2a)}}\cdot\vecc{k_1}}{s_1s_2\vecs{k_1}}, \nn
\end{eqnarray}
\begin{eqnarray}
s_1=-s\alpha_1=\tilde{s}_1+\vecs{k_1}-m^2,\quad s_2=
s\beta_1=M_2^2-m^2+\vecs{k_1}.
\end{eqnarray}
The quantities $M_{1,2}$ are the invariant masses of the jets. We
imply that the jet $X_1$ does not contain the detected pion. At
this point we need some model describing the jets. We use the
heavy fermion model, i.e., we consider the jet as a result of
heavy fermion decay. We assume the coupling constant of the
interaction of the pion with a nucleon and with heavy fermion to
be the same. We do not specify it as well as it is cancelled in
the asymmetry (\ref{eq:5}). So we have
\begin{eqnarray}
J^{(1a)}_\mu=\bar{u}(p_1')t^aO_\mu u(p_1),\quad
J^{(2a)}_\mu=\bar{u}(p_2')t^a\gamma_\mu u(p_2),\\\nn
O_\mu=\gamma_5\frac{\hat{p}_1-\hat{k}_1+M_1}{d_1}\gamma_\mu+
\gamma_\mu\frac{\hat{p}_1'+\hat{k}_1+m}{d_2}\gamma_5,\quad
\end{eqnarray}
$t^a$ are the generators of color group $SU(N)$ and
\begin{eqnarray} \label{eq:11}
d_1=\frac{1}{\beta\bar{\beta}}[\beta^2M_1^2+(\vecc{p}
+\beta\vecc{k_1})^2],\quad d_2=-\frac{m^2}{\beta}[\rho^2+\beta^2].
\nn
\end{eqnarray}
We use here the spin density matrices of the jets (i.e., heavy
fermions)
\begin{equation}
\sum_\lambda
u^\lambda(p_1')\bar{u}^\lambda(p_1')=\hat{p}_1'+M_1,\quad
\sum_\lambda u^\lambda(p_2')\bar{u}^\lambda(p_2')=\hat{p}_2'+M_2,
\end{equation}
It is important to note that we impose the gauge invariant form of
matrix element and after that we use the heavy fermion model. This
operations do not commute as well as the heavy fermion currents do
not satisfy the current conservation condition. It is a specific
of the considered model.

For the lowest order differential cross section we obtain
\begin{eqnarray} 2E_\pi\frac{d^3\sigma(p)}{d^3 p} = \frac{2\alpha_s^2}
{\pi^2\bar{\beta}}T_2\int\frac{d^2\vecc{k_1}}{\pi}
\frac{\Phi_{01}(\vecc{k_1},\vecc{p})\Phi_{02}(\vecc{k_1})}
{(\vecs{k_1})^2}=\frac{2\alpha_s^2}{\pi^2\bar{\beta}}T_2J^{(2)}_0,
\end{eqnarray}
with the explicit expressions for the impact factors $\Phi_{01}$,
$\Phi_{02}$ given in Appendix~A and $T_2=(N^2-1)/4$.
%The dimensionless quantities $m^4J^{(2)}_0(\rho)$ is presented in
%the Fig~2a.

The lowest order spin--dependent contribution to the cross section
arises from the interference of imaginary part of 1--loop
radiative correction (RC) of Feynman diagram (FD) Fig~1b,c with
the Born amplitude Fig 1a. We do not consider the RC from FD Fig.
1, believing that such kind FD contribute to the nucleon
resonances formation. Besides it do not contribute in the leading
logarithmical approximation (LLA)
\begin{equation}
z\sim 1,\quad\frac{\alpha_s}{\pi}\ll 1.
\end{equation}
We obtain in the lowest order
\begin{equation}
2E_\pi\frac{d\sigma}{{d^3\displaystyle p}}=
\frac{2\alpha_s^3}{\pi^2\bar{\beta}}RT_0I_0^{(3)},\;
\end{equation}
with
\begin{equation}
I_0^{(3)}=-\int\frac{d^2\vecc{k_1}}{\pi}\frac{d^2\vecc{k}}{\pi}
\frac{\Phi_{11}\Phi_{22}}{\vecs{k_1}
\vecc{k}^2(\vecc{k_1}-\vecc{k})^2},
\end{equation}
and the color factor
$$T_0=\big|Tr(t^at^bt^c)\big|^2=\frac{1}{16}\left[f_{abc}^2+
d_{abc}^2\right]=\frac{(N^2-1)(N^2-2)}{8N}.$$ The impact factors
$\Phi_{11,22}$ are given in Appendix B.

%The dependence of the dimensionless quantity $I_0^{(3)}m^4$ is given in
%Fig 2b.
Impact factors $\Phi_{11,22}$ contain the new mass parameters
$\tilde{M}_{1,2}$ which are intermediate jet state masses.

\section{Ladder expansion}

We had shown that the lowest order unpolarized and polarized cross
sections can be expressed in terms of impact factors of
projectiles moving in opposite directions which where introduced
first in the papers of H.~Cheng and T.~T.~Wu \cite{Cheng}. The
calculation of RC to them can be done following the method
developed by J.~Balitski and L.~N.~Lipatov \cite{BL}. It was shown
by these authors that in the LLA, the cross section has as well
the form of conversion of impact factors of colliding particles
with some universal kernel. Physically it corresponds to the
replacement of exchanged gluons by the reggeized gluons. The
reggeization states are taking into account in two factors. First
the Regge factor $(s/m^2)^{a(t)}$ must be introduced, where $a(t)$
is the Regge trajectory of gluon with the momentum squared $t$.
The second factor takes into account the contribution of inelastic
processes of emission of real gluons. These both contributions
suffer from the infrared divergences, however the total sum is
free of them.
\begin{figure}[htb]
\begin{center}
\includegraphics[angle=0,scale=.5]{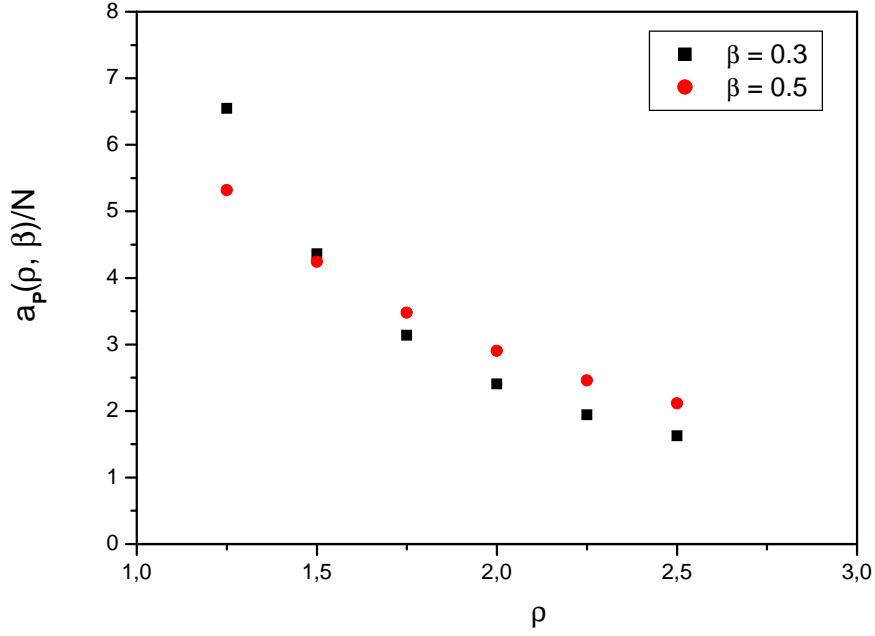}
\caption{The $\rho$, $\beta$ dependence of a partial contribution
to the pomeron intercept.} \label{pom}
\end{center}
\end{figure}
\begin{figure}[htb]
\begin{center}
\includegraphics[angle=0,scale=.5]{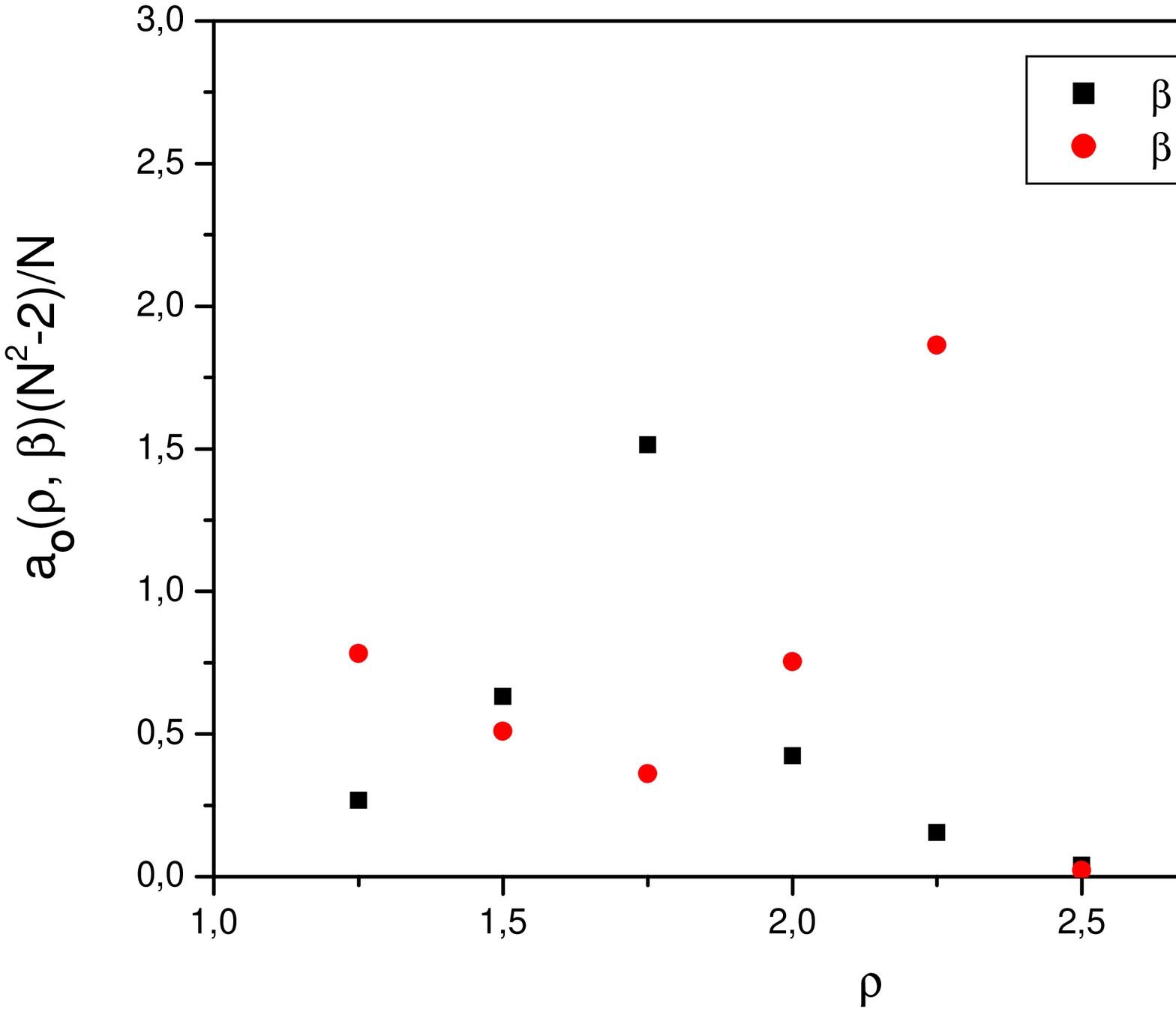}
\caption{The $\rho$, $\beta$ dependence of a partial contribution
to the odderon intercept.} \label{ode}
\end{center}
\end{figure}
For the RC to the unpolarized cross section we have
\begin{equation}
2E_\pi \frac{d\sigma}{d^3{\displaystyle p}}=
\frac{2\alpha_s^3}{\pi^3\bar{\beta}}T_3 \ln\frac{s}{m^2}J_1^{(2)},
\end{equation}
with color factor
$$T_3=Tr(t^at^b)Tr(t^{a'}t^{b'})(-f_{aa'd}f_{bdb'})=
\frac{N}{4}\left(N^2-1\right)$$ and
\begin{eqnarray} \label{eq:17}
J_1^{(2)}=2\int\frac{d^2\vecc{k_1}}{\pi}\frac{d^2\vecc{k'}}{\pi}
\frac{\Phi_1(\vecc{k_1},\vecc{p})}{\vecs{k_1}(\vecc{k_1}-\vecc{k'})^2}
\left[\frac{\Phi_2(\vecc{k'})}{\vecs{k'}}-
\frac{\Phi_2(\vecc{k_1})}{\vecs{k'}+(\vecc{k_1}-\vecc{k'})^2}\right],
\end{eqnarray}
with $\Phi_1$ and $\Phi_2$ given in Appendix~C.

This formula can be inferred from the result for the non forward
high energy scattering amplitude obtained in the paper \cite{BL}
\begin{equation}
I_1=3is\frac{\alpha_s}{2}\ln(s/m^2)\int\frac{d^2\vecc{k}}{\pi}
\frac{\Phi^{AA'}(\vecc{k},\vecc{q})}
{\vecc{k}^2(\vecc{q}-\vecc{k})^2}
\Phi^{BB'}*K(\vecc{k},\vecc{k'},\vecc{q}),
\end{equation}
with definition
\begin{align}
&\Phi^{BB'}*K(\vecc{k},\vecc{k'},\vecc{q})= \nn\\ &\quad
\int\frac{d^2\vecc{k'}}{\pi}\Bigg\{\bigg[-\vecc{q}^2+\frac{
\vecc{k}^2(\vecc{q}-\vecc{k'})^2+\vecs{k'}(\vecc{q}-\vecc{k})^2}{(
\vecc{k}-\vecc{k}')^2}\bigg]\frac{\Phi^{BB'}(\vecc{k'},\vecc{q})}
{\vecs{k'}(\vecc{q}-\vecc{k'})^2}\\ \nn &\quad -
\frac{\Phi^{BB'}(\vecc{k},\vecc{q})}{(\vecc{k}-\vecc{k'})^2}
\bigg[\frac{\vecc{k}^2}{\vecs{k'}+(\vecc{k}-\vecc{k'})^2}+
\frac{(\vecc{q}-\vecc{k})^2}{(\vecc{q}-\vecc{k'})^2+(\vecc{k}-
\vecc{k'})^2}\bigg]\Bigg\}.
\end{align}
The formula (\ref{eq:17}) can be obtained from the last general
one by putting $\vecc{q}=0$. Note that in the formula obtained in
\cite{BL} the used color group was $SU(2)$.

Let now consider the LLA RC to the polarized part of the
differential cross section. There are presented three types of
contributions corresponding to three different choices of two
gluons which are involved in the reggeization procedure in the
lowest order RC
\begin{equation}
2E_\pi\frac{d\sigma}{d^3{\displaystyle p}}=
\frac{\alpha_s^4}{\pi^3}RT_1I_1^{(3)}\ln\frac{s}{m^2},\quad
I_1^{(3)}=\int\frac{d^2\vecc{k}}{\pi}\frac{d^2\vecc{k'}}{\pi}
\frac{d^2\vecc{k_1}}{\pi} \left[I_{12}+I_{13}+I_{23}\right],
\end{equation}
with the color factor
$$T_1=Tr t^at^bt^c Tr t^{b'}t^{a'}t^c f_{aa'd}f_{bdb'}=
\frac{(N^2-1)}{8}.$$

Symmetry reasons lead to the conclusion that $I_{13}$ and $I_{23}$
contributions are equal to $I_{12}$ one. For $I_{12}$ we have
\begin{align} \label{eq:21}
I_{12}&=\frac{\Phi_1^{(12)}(\vecc{k},\vecc{k_1})}{\vecc{k}^2
(\vecc{k_1}-\vecc{k})^2}
\left\{\frac{\Phi_2^{(12)}(\vecc{k'},\vecc{k_1})}{\vecs{k'}
(\vecc{k_1}-\vecc{k'})^2}\left[-\vecs{k_1}+
\frac{\vecc{k}^2(\vecc{k_1}-\vecc{k'})^2+\vecs{k'}
(\vecc{k_1}-\vecc{k})^2}{(\vecc{k}-\vecc{k'})^2}\right]\right. \nn \\
&\quad -\left.
\frac{\Phi_2^{(12)}(\vecc{k},\vecc{k_1})}{(\vecc{k}-\vecc{k'})^2}
\left[\frac{\vecc{k}^2}{\vecs{k'}+(\vecc{k}-\vecc{k'})^2}+
\frac{(\vecc{k_1}-\vecc{k})^2}{(\vecc{k_1}-\vecc{k'})^2+
(\vecc{k}-\vecc{k'})^2}\right]\right\}.
\end{align}
%For $I_{13}$ we have
%\begin{eqnarray}
%I_{13}&=&\frac{\Phi_1^{(13)}(\vecc{k},\vecc{k}-\vecc{k_1})}
%{\vecc{k}^2\vecs{k_1}}\left\{\frac{\Phi_2^{(13)}(\vecc{k'},\vecc{k}-
%\vecc{k_1})}{\vecs{k'}(\vecc{k}-\vecc{k_1}-\vecc{k'})^2}\right.\\\nn
%&\times&\left[\frac{\vecc{k}^2(\vecc{k}-\vecc{k_1}-
%\vecc{k'})^2+\vecs{k'}\vecs{k_1}} {(\vecc{k}-\vecc{k'})^2}-
%(\vecc{k}-\vecc{k_1})^2\right]\\ &-&\left.
%\frac{\Phi_2^{(13)}(\vecc{k},\vecc{k}-\vecc{k_1})}{(\vecc{k}-
%\vecc{k'})^2}\left[\frac{\vecc{k}^2}{\vecs{k'}+(\vecc{k}-
%\vecc{k'})^2}+\frac{\vecs{k_1}}{(\vecc{k}-\vecc{k_1}-\vecc{k'})^2+
%(\vecc{k}-\vecc{k'})^2}\right]\right\}. \nn
%\end{eqnarray}
%For $I_{23}$ we have
%\begin{eqnarray}
%I_{23}&=&\frac{\Phi_1^{(23)}(\vecc{k_1}-\vecc{k},-\vecc{k})}
%{\vecs{k_1}(\vecc{k_1}-\vecc{k})^2}\left[\frac{\Phi_2^{(23)}
%(\vecc{k_1}-\vecc{k'},-\vecc{k})}{(\vecc{k_1}-\vecc{k'})^2
%(\vecc{k}+\vecc{k_1}-\vecc{k'})^2}\right.\\ \nn&\times&
%\left[-\vecc{k}^2+\frac{(\vecc{k_1}-\vecc{k})^2(\vecc{k}+\vecc{k_1}
%-\vecc{k'})^2+(\vecc{k_1}-\vecc{k'})^2\vecs{k_1}}
%{(\vecc{k}-\vecc{k'})^2}\right] \\\nn
%&-&\left.\frac{\Phi_2^{(23)}(\vecc{k_1}-\vecc{k},-\vecc{k})}
%{(\vecc{k}-\vecc{k'})^2}\left[\frac{(\vecc{k_1}-\vecc{k})^2}
%{(\vecc{k_1}-\vecc{k'})^2+(\vecc{k}-\vecc{k'})^2}+\frac{\vecs{k_1}}
%{(\vecc{k}-\vecc{k'})^2+(\vecc{k_1}-\vecc{k'}+\vecc{k})^2}
%\right]\right].
%\end{eqnarray}
The details of impact factor calculations are given in the
Appendices A--D. In the Appendix E we give some details used for
performing the integration over the transversal component of the
loop momenta.

%The dependence of the function $m^4I_1^{(3)}$ on $\rho$ at fixed
%$\beta$ is given in Fig.2.

\section{Discussion}

The quantities
\begin{eqnarray}
a_\Ode(\rho,\beta)=\frac{N}{(N^2-2)}\frac{I_1^{(3)}}{I_0^{(3)}},
\quad a_\Pom(\rho,\beta)=N\frac{J_1^{(2)}}{J_0^{(2)}},
\end{eqnarray}
may be interpreted as a partial contributions to the odderon and
pomeron intercepts. Their dependence on $\rho$, $\beta$ is
illustrated in Fig.~\ref{pom} and Fig.~\ref{ode}. The jet's masses
was supposed to be larger than 1\,GeV.

The size of contributions to the polarized and unpolarized
differential cross sections depends on the used jet model as well
as on the choice of the vertices which describe the transition of
the nucleon to the jet and on the choice of the vertex function
which describes the conversion of one sort of jet to the jet of
another sort. Because of the here used choice
$V_\mu(k)=\gamma_\mu$, we are forced to put on the gauge
conditions. Another possible choice,
$V_\mu(k)=[\gamma_\mu,\hat{k}]/M$, leads to the zero contribution
to the asymmetry (in the limit of infinite large $s$).

We see that one-spin effect is rather large. Another mechanisms of
one-spin asymmetry associated with the nucleon resonances in
intermediate state and final state interaction in $ep\to ep\pi$
process was considered in papers \cite{AVKN,AAKR} where the effect
was of the same order.

\section*{Acknowledgments}

The work was supported by INTAS-00366. E.~Z. is gratefull to RFBR
for the grant 0001-00617.

\newpage
\section*{Appendix A}
\renewcommand{\theequation}{A.\arabic{equation}}
\setcounter{equation}{0}

The explicit expressions for the lowest order impact factors in
the unpolarized case are
\begin{align}
\Phi_{01}(\vecc{k_1},\vecc{p})&=\frac{1}{4s^2}
Sp(\hat{p}_1'+m)O_\mu(\hat{p}_1+m)
\tilde{O}_\nu p_{2}^\mu p_{2}^\nu,\nn\\
\Phi_{02}(\vecc{k_1})&=\frac{1}{4s_2^2}Sp(\hat{p}_2'+M_2)\gamma_\mu
(\hat{p}_2+m)\gamma_\nu k_{1\bot}^\mu k_{1\bot}^\nu.
\end{align}
To simplify the calculation of the traces we write down here the
useful relations
\begin{gather} \label{eq:a2}
p_1'=\frac{a_1}{s}q_2+\bar{\beta}q_1-p_\bot-k_{1\bot},\quad
p_1'+k_1=p_1-\frac{\vecc{p}^2}{s}q_2-\beta q_1-p_\bot,\nn\\
p_1-k_1=p_1+\frac{s_1}{s}q_2-k_{1\bot},\quad
p_2'=p_2+\frac{s_2}{s}q_1+k_{1\bot},\nn\\
a_1=(M_1^2+(\vecc{p}+\vecc{k_1})^2)/\bar{\beta},
\end{gather}
which follow from the  on mass shell conditions. Explicit
expressions for $\Phi_{01}$ and $\Phi_{02}$ are
\begin{align}
\Phi_{01}(\vecc{k_1},\vecc{p})&=\frac{\beta^4\bar{\beta}\vecs{k_1}}
{2\Big[\beta^2m^2+\vecc{p}^2\Big]\Big[\beta^2m^2+(\vecc{p}-\vecc{k_1}
\beta)^2\Big]}, \nn \\\Phi_{02}\vecc{k_1})&=\frac{2\vecc{k_1}^2\Big[
\vecs{k_1}+(M_2-m)^2\Big]}{\Big(\vecc{q}^2+M_2^2-m^2\Big)^2}
\end{align}

\section*{Appendix B}
\renewcommand{\theequation}{B.\arabic{equation}}
\setcounter{equation}{0}

The lowest order contribution to the impact factor corresponding
to the polarized proton with 4-momentum $p_1$ is
\begin{equation}
\Phi_{11}=(1/\bar{\beta})\Phi_1^b+\Phi_1^c.
\end{equation}
The quantity $\Phi_1^b$ has a form
\begin{align}
\Phi_1^b&=\frac{1}{4s^3R}
Sp(\hat{p}_1+m)(-\gamma_5\hat{a})\tilde{O}_\mu p_2^\mu
(\hat{p}_1'+m)\hat{p}_2 \\\nn&\quad \times
(\hat{p}_1'+\hat{k}_1-\hat{k}+m)O_{1\nu}p_2^\nu.
\end{align}
The Sudakov decomposition of the 4-vectors $p_1'$ is given in
(A.2). The exchanged gluon expansions are
\begin{gather}
k=\alpha q_2+k_\bot,\quad s\alpha=\vecc{k}^2-d_3,\\ \nn
k_1=\alpha_1 q_2+k_{1\bot},\quad s\alpha_1=\vecs{k_1}-d_1,
\end{gather}
the quantity $O_{1\mu}$ is equal
$$
O_{1\mu}=(1/d_2)\gamma_\mu(\hat{p}_1'+\hat{k}_1+m)\gamma_5+(1/d_3)\gamma_5(
\hat{p}_1-\hat{p}_1+m)\gamma_\mu,
$$
with $d_{1,2}$ given in (\ref{eq:11}) and
$$
d_3=\frac{1}{\beta\bar{\beta}}\left[\beta^2m^2+(\vecc{p}
+\beta\vecc{k})^2\right].
$$
The quantity $\Phi_1^c$ has a form
\begin{gather}
\Phi_1^c=\frac{1}{4s^3R}Sp(\hat{p}_1+m)(-\gamma_5\hat{a})
\tilde{O}_\mu p_2^\mu(\hat{p}_1'+m)O_{2\nu}p_2^\nu
(\hat{p}_1-\hat{k}+m)\hat{p}_2,\nn\\
O_{2\mu}=\gamma_\mu\frac{\hat{p}_1'+\hat{k}_1-\hat{k}+m}{d_4}
\gamma_5+ \gamma_5\frac{\hat{p}_1-\hat{k}+m}{d_1}\gamma_\mu,
\end{gather}
with
$$
d_4=-\frac{1}{\beta\bar{\beta}}\left[\beta^2m^2+
(\vecc{p}+\beta\vecc{k})^2\right]
$$
and with the same expression for $p_1'$.

The relevant representation for the exchanged gluon 4--momenta is
following
\begin{gather}
k=\alpha q_2+k_\bot, \quad s\alpha=\vecc{k}^2,
\\ \nn k_1=\alpha_1 q_2+k_{1\bot},\quad
s\alpha_1=\vecs{k_1}-d_1
\end{gather}
and the same expression for $p_1'$.

Impact factor for the unpolarized proton $\Phi_2$ has a form
\begin{align}
\Phi_{22}&=\frac{1}{s\beta s\beta_1 s(\beta-\beta_1)}
\frac{1}{4}Sp(\hat{p}_2+m)\hat{k}_{1\bot}\nn\\
&\quad \times(\hat{p}_2+\hat{k}_1+M_2)(\hat{k}_{1\bot}-
\hat{k}_\bot)(\hat{p}_2+\hat{k}+\tilde{M}_2)\hat{k}_\bot.
\end{align}
For then calculation of the trace for $\Phi_2$ we used Sudakov
representation
\begin{eqnarray} &k=\beta q_1+k_\bot,\quad s\beta=
\tilde{M}_2^2-m^2+\vecc{k}^2,&\nn\\
&k_1=\beta_1 q_1+k_{1\bot},\quad s\beta_1=M_2^2-m^2+\vecs{k_1}.&
\end{eqnarray}

\section*{Appendix C}
\renewcommand{\theequation}{C.\arabic{equation}}
\setcounter{equation}{0}

The impact factors for the case of unpolarized protons are
\begin{gather}
\Phi_1(\vecc{k_1},\vecc{p})=\Phi_{01}(\vecc{k_1},\vecc{p}),\\ \nn
\Phi_2(\vecc{k'})=\frac{1}{4ss_2}Sp(\hat{p}_2+\hat{k}_\bot'+
\beta'q_1+M_2)\hat{p}_1(\hat{p}_2+m)\hat{k}'_\bot
\end{gather}
where
$$ s_2=M_2^2-m^2+\vecs{k'}.$$

\section*{Appendix D}
\renewcommand{\theequation}{D.\arabic{equation}}
\setcounter{equation}{0}

Let us now give the expressions for the impact factors in the case
of RC to the polarized cross sections. For the $\Phi_1^{(12)}$ in
(\ref{eq:21}) we have
\begin{equation}
\Phi_1^{(12)}(\vecc{k},\vecc{k_1})=\Phi_{1b}^{(12)}+\Phi_{1c}^{(12)},
\end{equation}
with
\begin{align}
\Phi_{1b}^{(12)}&=\frac{1}{4sRs_{11}s_{12}}Sp(\hat{p}_1+m)
(-\gamma_5\hat{a})\tilde{O}_\mu p_2^\mu(\hat{p}'_1+m)\\ \nn
&\quad{\times}(\hat{k}_{1\bot}-\hat{k}_\bot)
(\hat{p}_1'+\hat{k}_1-\hat{k}+m)O_{1\nu}k_\bot^\nu ;
\end{align}
and
\begin{align}
\Phi_{1c}^{(12)}&=\frac{1}{4sRs_{21}s_{22}}Sp(\hat{p}_1+m)
(-\gamma_5\hat{a})\tilde{O}_\mu p_2^\mu(\hat{p}'_1+m)\\ \nn
&\quad{\times}O_{2\mu}(k_{1\bot}-k_\bot)^\mu
(\hat{p}_1-\hat{k}+m)\hat{k}_\bot,
\end{align}
with the substitutions similar to ones for $\Phi_1^b$ from
Appendix B for the momenta including $\Phi_{1b}^{(12)}$ and the
substitutions similar to ones for $\Phi_1^c$ for
$\Phi^{(12)}_{1c}$. Besides
\begin{gather}
s_{11}=(\vecc{k_1}+\vecc{p})^2-(\vecc{k}+ \vecc{p})^2,\quad
s_{12}=m^2-\tilde{s}_1(m^2)-\vecc{k}^2,\\ \nn
s_{21}=-\vecc{k}^2,\quad
s_{22}=m^2-\tilde{s}_1+\vecc{k}^2-\vecs{k_1}.
\end{gather}
The impact factor of unpolarized proton in this case have a form
\begin{eqnarray}
\Phi_2^{(12)}(\vecc{k'},\vecc{k_1})&=&\frac{1}{4s_2s_{23}s_{13}}
Sp(\hat{p}_2+m)\hat{k}_{1\bot}(\hat{p}_2+\hat{k}_{1\bot}+\beta_1
\hat{q}_1+M_2) \\ \nn &{\times}&
(\hat{k}_{1\bot}-\hat{k}'_\bot)(\hat{p}_2+\hat{k}'_\bot+
\beta'\hat{q}_1+\tilde{M}_2)\hat{k}'_\bot, \nn
\end{eqnarray}
with
\begin{eqnarray}
s_{13}=\tilde{M}_2^2-M_2^2-\vecs{k_1}+\vecs{k'},\quad
s_{23}=\tilde{M}_2^2-m^2+\vecs{k'},
\end{eqnarray}
and
\begin{eqnarray}
s\beta_1=M_2^2-m^2+\vecs{k_1}, \quad
s\beta'=\tilde{M}_2^2-m^2+\vecs{k'}.
\end{eqnarray}

\section*{Appendix E}
\renewcommand{\theequation}{E.\arabic{equation}}
\setcounter{equation}{0}

Here we give some details used in performing the loop momenta
integration.

When calculating the relevant trace we use the Shouthen identity
$$
(p_1p_2p_3p_4)Q_\mu=(\mu p_2p_3p_4)Qp_1+(p_1\mu p_3p_4)Qp_2+(p_1p_2\mu p_4)Qp_3+
(p_1p_2p_3\mu)Qp_4.
$$
This identity permits to express all conversions with Levi-Chivita
tensor in the standard form. For instance
$$
(p_1p_2k_1p_\bot)(ak_2)=(p_1p_2ap_\bot)(k_1k_2)+
(p_1p_2k_1a)(p_\bot k_2).
$$
The second term in the right side of this equation can be
expressed through the
\begin{equation}
E=(p_1p_2ap)=\frac{s}{2}m \rho R,
\end{equation}
using the relation
\begin{gather}
\displaystyle\int\frac{d^2\vecc{k}}{\pi}\frac{d^2\vecc{k_1}}{\pi}
\frac{d^2\vecc{k'}}{\pi}F(\vecc{k},\vecc{k_1},\vecc{k'},\vecc{p})
\vecc{k_i}=\frac{\vecc{p_i}}{\vecc{p}^2}
\displaystyle\int\frac{d^2\vecc{k}}{\pi}\frac{d^2\vecc{k_1}}{\pi}
\frac{d^2\vecc{k'}}{\pi}F(\vecc{k},\vecc{k_1},\vecc{k'},
\vecc{p})\vecc{p}.\vecc{k}.
\end{gather}

\end{document}